\pdfoutput=1
\documentclass[nofootinbib,unsortedaddress,superscriptaddress]{revtex4}
\usepackage{amsmath}
\usepackage{amssymb}
\usepackage{graphicx}
\usepackage{subfig}

\makeatother
\date{\today}
\newcommand{\typevector}[1]{\boldsymbol #1}
\newcommand{\vecr}{\typevector r}

\newcommand{\vecx}{\typevector x}
\newcommand{\vecy}{\typevector y}
\newcommand{\vecz}{\typevector z}

\begin{document}
\title{Replicating Kernels with a Short Stride Allows Sparse Reconstructions with Fewer Independent Kernels}

\author{Peter F. Schultz}
\email[Email address: ]{pschultz@newmexicoconsortium.org}
\affiliation{New Mexico Consortium}
\author{Dylan M. Paiton}
\affiliation{University of California, Berkeley}
\author{Wei Lu}
\affiliation{University of Michigan}
\author{Garrett T. Kenyon}
\affiliation{Los Alamos National Laboratory and New Mexico Consortium}

\begin{abstract}
In sparse coding it is common to tile an image into nonoverlapping patches, and then use a dictionary to create a sparse representation of each tile independently. In this situation, the overcompleteness of the dictionary is the number of dictionary elements divided by the patch size. In deconvolutional neural networks (DCNs), dictionaries learned on nonoverlapping tiles are replaced by a family of convolution kernels.  Hence adjacent points in the feature maps (V1 layers) have receptive fields in the image that are translations of each other. The translational distance is determined by the dimensions of V1 in comparison to the dimensions of the image space.  We refer to this translational distance as the stride.

We implement a type of DCN using a modified Locally Competitive Algorithm (LCA) to investigate the relationship between the number of kernels, the stride, the receptive field size, and the quality of reconstruction. We find, for example, that for 16x16-pixel receptive fields, using eight kernels and a stride of 2 leads to sparse reconstructions of comparable quality as using 512 kernels and a stride of 16 (the nonoverlapping case). We also find that for a given stride and number of kernels, the patch size does not significantly affect reconstruction quality. Instead, the learned convolution kernels have a natural support radius independent of the patch size.
\end{abstract}

\maketitle

\section*{Introduction}
\label{intro}
Sparse coding has been widely used to model the structure of of images.  Typically a good sparse code requires learning an overcomplete dictionary of weights~\cite{Olshausen2013}.  For image patches as small as 16-by-16 pixels, an overcomplete dictionary will still require hundreds to thousands of dictionary elements.  It therefore would be advantageous to identify techniques that would reduce the number of independent weights that have to be learned.

For natural images, we expect image statistics to be similar at different parts of the image, and that important features of the image will be localized in space.  This motivates the idea of deconvolutional networks~\cite{ZeilerKrishnanTaylorFergus2010}, where the image is modeled as a sum of convolutions.  Each kernel appearing in this sum captures a particular local image feature (for example, a Gabor filter), and is applied to a family of image patches, all of the same size but shifted relative to each other, by an amount we refer to as the stride. 

Using deconvolutional networks can significantly reduce the number of free parameters used in learning the weights.  This reduction occurs because the number of image patches that cover a given image pixel is greater than the number of kernels, due to the overlap in patches arising from the convolution.  Hence fewer independent kernels are needed to achieve the same coverage of the image. The smaller the stride, the larger the effect of overlapping patches will be.  In this work, we explore the relation between the stride, the number of convolution kernels, and the quality of the resulting image reconstructions.

\section*{Background}
\label{background}
In sparse coding, we seek to approximate an $M\times N$ image~${\vecx}$ in the form
\begin{equation}
x\approx\sum_{k=1}^K y_k\phi_k\;,
\label{sparsecoding}
\end{equation}
where $\phi_1,\ldots,\phi_k$ are $M\times N$-sized dictionary elements and $y_1,\ldots,y_k$ are scalars that constitute the representation of~$\vecx$.  By considering the $y_k$ to be elements of a vector~$\vecy$ and the $\phi_k$ to be column vectors of a $(M\times N)$-by-$K$ matrix~$\Phi$, we can write
\begin{equation}
\vecx\approx \Phi \vecy.
\end{equation}
In a sparse representation, only a small fraction of the~$y_k$ are nonzero.  If $K=MN$ there is typically a unique $\vecy$ such that $\Phi \vecy=\vecx$; however, this representation is unlikely to be sparse.  When $K=MN$, close approximations are typically not sparse and sparse representations are typically not good approximations.  If $K<MN$, the undercomplete case, it may be possible to find a close approximation, but it is even more unlikely that this approximation will be sparse.  If $K>MN$, the overcomplete case, there are typically infinitely many solutions~$\vecy$ to the exact equation $\Phi\vecy = \vecx$, and we can expect that there are sparse representations~$\vecy$ that are good approximations to~$\vecx$.  The ratio $K/MN$ is the overcompleteness factor.  Typically, researchers have used dictionaries in the range of 0.75 to~10 times overcomplete \cite{Olshausen2013}, \cite{OlshausenField1996}, \cite{OlshausenField1997}, \cite{RozellJohnsonBaraniukOlshausen2008}, \cite{ZeilerKrishnanTaylorFergus2010}, \cite{ZylberbergMurphyDeWeese2011}.

There are many ways to balance the closeness of the approximation with the sparseness of the representation.  In this work, we will minimize an energy function
\begin{equation}
  E(\vecy) = \frac12\|\vecx-\Phi \vecy\|_2^2 + C(\vecy),
  \label{sparsecodingenergy}
\end{equation}
where $C(\vecy)$ is a cost function that penalizes nonsparse vectors~$\vecy$.  Here, we will use an approximation to the $l^0$ norm using a threshold parameter~$\lambda$:
\begin{equation}
   C(\vecy) = \sum_k C_{\lambda}(y_k),\;\hbox{with}\; C_\lambda(a) = \begin{cases} \lambda & \hbox{if}\;|a|\geq \lambda \\ 0& \text{if}\;|a|<\lambda
\label{l0cost}
\end{cases}
\end{equation}

Zeiler et al. introduced deconvolutional networks~\cite{ZeilerKrishnanTaylorFergus2010}, in which we approximate $\vecx$ by finding feature maps $\vecz_j$ convolved with convolutional kernels $f_j$:
\begin{equation}
    \vecx\approx\sum_{j=1}^J f_j * \vecz_j
\end{equation}
The kernels $f_j$ are small patches (for example, Gabor filters), and the $\vecz_j$ are of size $M\times N$, with adjustments for edge effects.

The task is then to minimize the energy function
\begin{equation}
  E(z) = \frac12\|x-\sum_j f_j * \vecz_j\|_2^2 + C(\vecz).
  \label{deconvenergy}
\end{equation}

\begin{figure}
\unitlength=0.01in
\begin{picture}(550,300)
\put(0,0){\makebox(0,0)[lb]{\includegraphics[height=2.75in]{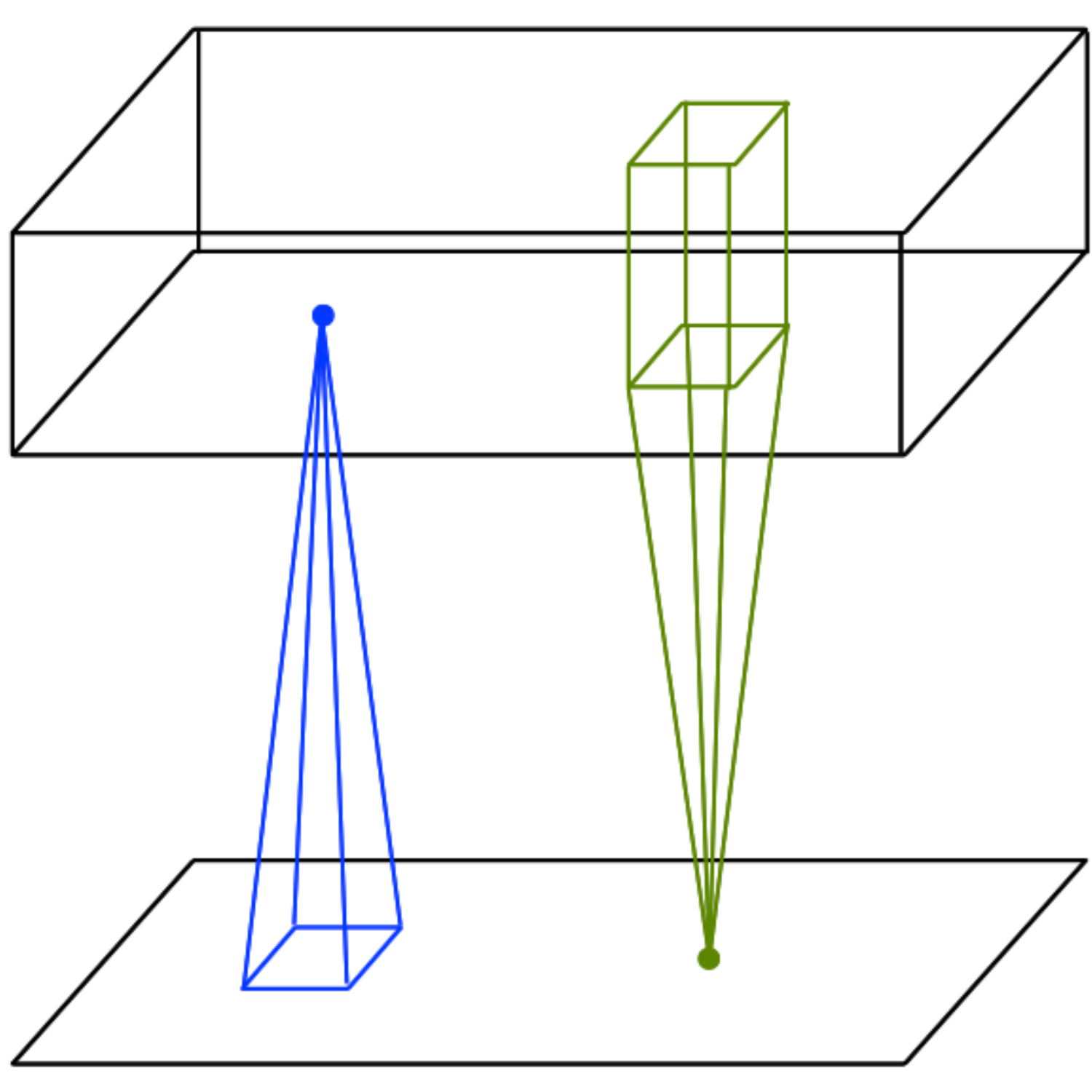}}}
\put(400,40){\makebox(0,0){Image $\vecx$}}
\put(400,230){\makebox(0,0){V1 layer $\vecz$}}
\put(265,185){\makebox(0,0){$M$}}
\put(150,280){\makebox(0,0){$N$}}
\put(285,240){\makebox(0,0){$J$}}
\put(265,30){\makebox(0,0){$M$}}
\put(150,70){\makebox(0,0){$N$}}
\end{picture}
\caption{The image layer~$\vecx$ and the V1 layer~$\vecz$.  Blue shows the receptive field in~$\vecx$ from a single point in~$\vecx$.  Green shows for a point in~$x$ the region in~$\vecz$ whose receptive field contains the point.}
\label{featuremapcube}
\end{figure}
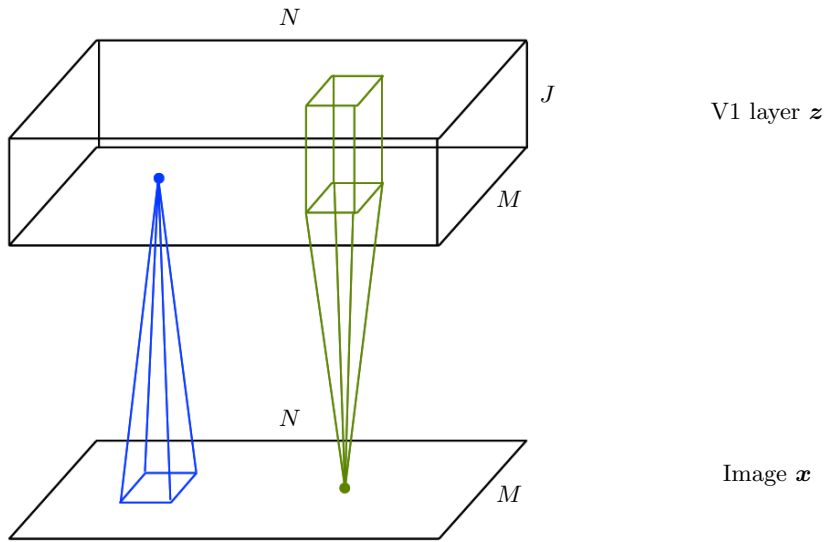

It is natural to consider the collection of feature maps as an $M$-by-$N$-by-$J$ grid above the $M$-by-$N$ image layer.  Each gridpoint has a receptive field in the image layer~$\vecx$, centered at the point below the voxel, and whose size is the same as the patch size of the $f_j$.  Similarly, each pixel in the image influences a size($f_j$)-by-$J$ region of the feature maps (figure~\ref{featuremapcube}).  Thus, the feature maps play a similar role to the V1~layer in biological vision systems.  Note that moving the blue point in the V1~layer by one pixel will move its receptive field by 1~pixel in $\vecx$-space; that is, the network has a stride of~1.

The $f_j$ are analogous to the $\phi$ in equation~(\ref{sparsecodingenergy}) and the $\vecz_j$ are analogous to the~$\vecy$.  We note that deconvolutional networks can be described in terms of the formalism of approximation~(\ref{sparsecoding}).  Namely, there are $K=JMN$ dictionary elements:  for each of the $MN$ pixels, translate each of the $J$ convolutional kernels to be centered at that pixel.  Accordingly, the overcompleteness factor is $K/MN = J$.  Note that the overcompleteness does not depend on the patch size of the $f_j$.

Deconvolutional networks can also be extended to the case where the feature maps are downsampled from the original image size.  In this case, the number of dictionary elements is determined by the number of pixels in the domain of $\vecz_j$. For example, if the $\vecz_j$ are $M/2$-by-$N/2$, the number of dictionary elements is $J\cdot(M/2)\cdot(N/2)$ and the overcompleteness factor is $J/4$.  Again, this factor does not depend on the patch size.  We visualize the $J$ feature maps as an $(M/2)$-by-$(N/2)$-by-$J$ grid, but with a lower density of points than in the image layer, so that each gridpoint of the feature map lies above the center of its receptive field.  Because of this lower density, moving one pixel in $\vecz$-space shifts the receptive field by 2~pixels in $\vecx$-space, giving a stride of 2.
In general, feature maps with dimensions $(M/F)$-by-$(N/F)$ will have a stride of~$F$.

For a deconvolutional network, each neuron is directly affected only by nearby neurons.  Accordingly, we can use Locally Competitive Algorithms (LCA) of Rozell et al.~\cite{RozellJohnsonBaraniukOlshausen2008}  For a two-layer network consisting of an image~$\vecx$ and a V1-type layer~$\vecy$, the LCA dynamics are as follows:
\begin{align}
   \dfrac{du_k}{dt} &= -u_k + (\Phi^T\vecx)_k - (\Phi^T\Phi \vecy-\vecy)_k, \label{lcadynamics}\\
   y_k &= T(u_k).\label{lcatransfer}
\end{align}
Here $u_k$ is an internal state variable corresponding to V1-neuron $y_k$, and $T(u)$ is a transfer function.  For the thresholded $l^0$ cost function in equation~(\ref{l0cost}), $T(u)$ is the hard threshold function
\begin{equation}
   T(u) = \begin{cases} u & \text{if}\;u\geq\lambda \\ 0 & \text{if}\; u<\lambda. \end{cases}
\end{equation}
Equation~(\ref{lcadynamics}) has an elegant motivation in terms of leaky integrators.  The $-u_k$ term provides leakiness of the internal state, the $(\Phi^T \vecx)_k$ term charges the neuron up based on input, and the $\Phi^T\Phi \vecy-\vecy$ term provides local competition between different $y$-neurons.  If the columns of $\Phi$ have unit $l^2$ norm, the $-\vecy$ part eliminates self-interactions.  This system will converge to a local minimum of the energy function~(\ref{sparsecodingenergy}).  One advantage of LCAs is that the selection of local minimum is more stable as $\vecx$ varies continuously, as in video.  Also, the local nature of LCAs and DCNs means that the algorithm is well-suited to being implemented in hardware based, for example, on FPGAs or memristor arrays.

Note that in equation~(\ref{lcadynamics}), we can group the two terms involving $\Phi^T$; thus we see that the input~$\vecx$ enters the equation only as part of the residual, $\vecx-\Phi \vecy$, which is the part of the input that has not yet been accounted for by the representation~$\vecy$.  Our implementation therefore introduces an intermediate ``residual layer'' which holds the value $\vecr = \vecx-\Phi \vecy$.   Our implementation is therefore given by
\begin{align}
r_j &= x_j - (\Phi \vecy)_j\\
\dfrac{du_k}{dt} &= -u_k + y_k + (\Phi^T\vecr)_k\label{lcadynamicswithresidual}\\
   y_k &= T(u_k).
\end{align}
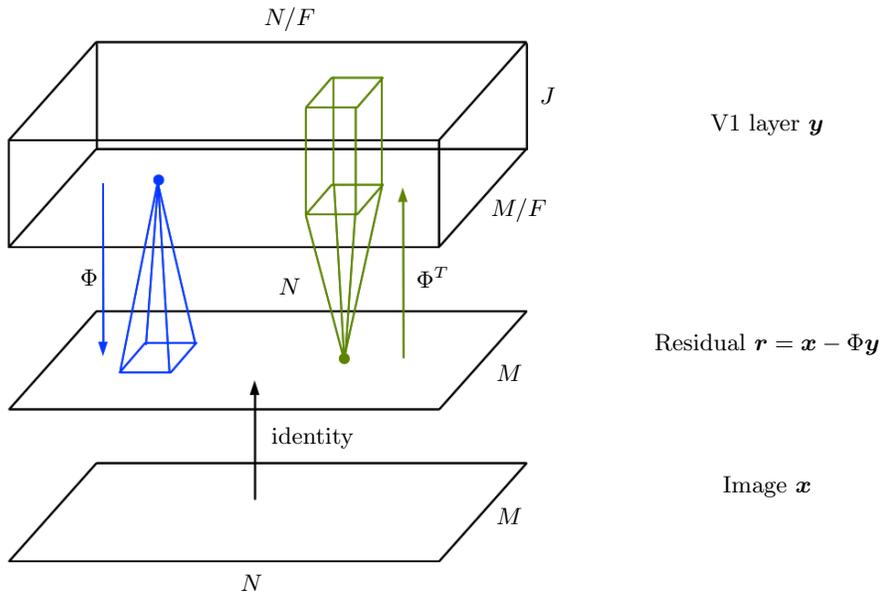
\begin{figure}
\unitlength=0.01in
\begin{picture}(550,300)(0,-10)
\put(0,0){\makebox(0,0)[lb]{\includegraphics[height=2.75in]{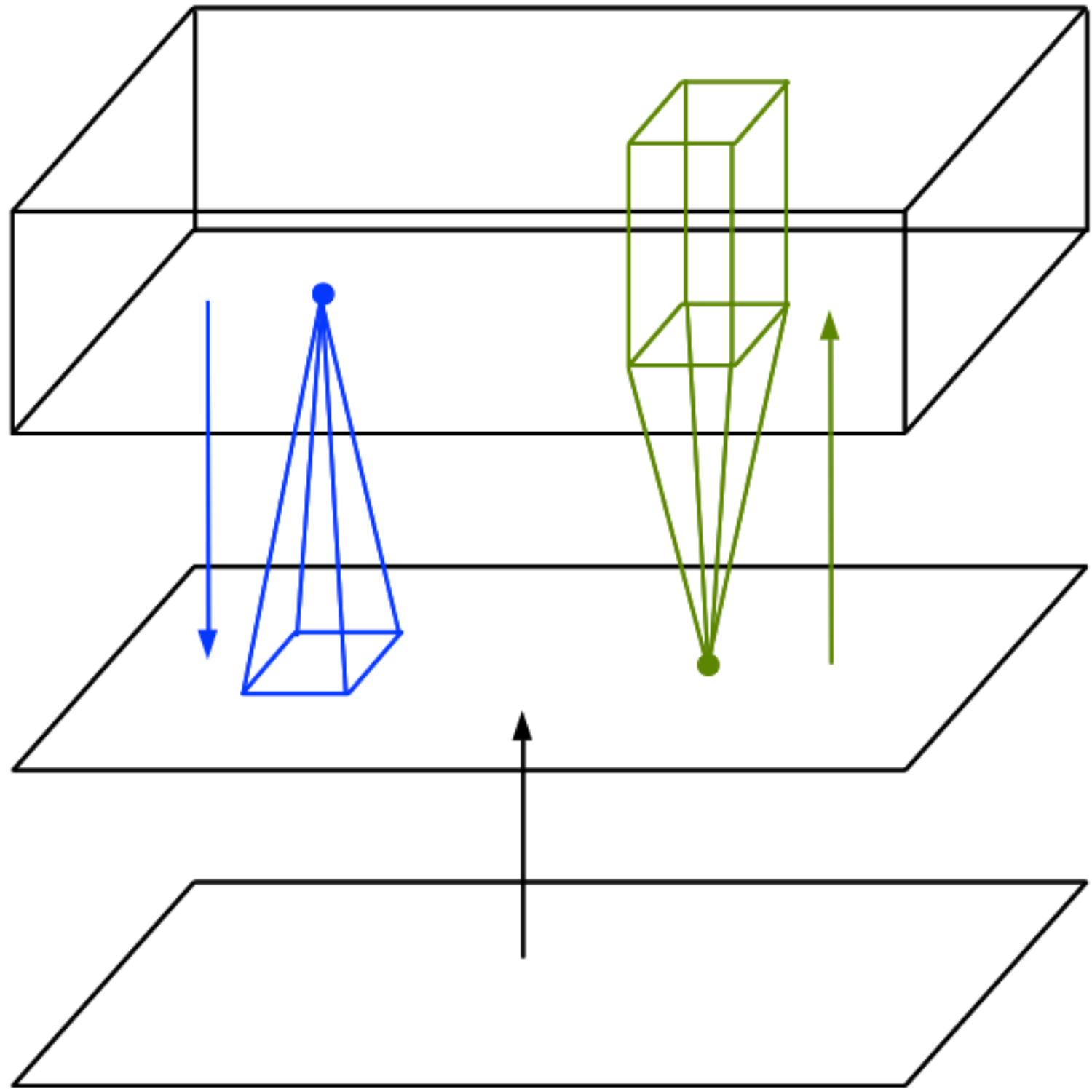}}}
\put(400,40){\makebox(0,0){Image $\vecx$}}
\put(400,115){\makebox(0,0){Residual $\vecr = \vecx - \Phi\vecy$}}
\put(400,230){\makebox(0,0){V1 layer $\vecy$}}
\put(270,185){\makebox(0,0){$M/F$}}
\put(150,285){\makebox(0,0){$N/F$}}
\put(285,245){\makebox(0,0){$J$}}
\put(45,150){\makebox(0,0){$\Phi$}}
\put(225,150){\makebox(0,0){$\Phi^T$}}
\put(265,100){\makebox(0,0){$M$}}
\put(150,145){\makebox(0,0){$N$}}
\put(162,66){\makebox(0,0){identity}}
\put(265,25){\makebox(0,0){$M$}}
\put(130,-10){\makebox(0,0){$N$}}
\end{picture}
\caption{Successful predictions by the V1 representation inhibits the residual layer by means of~$\Phi$.  Nonzero values in the residual layer excite the V1 representation by means of~$\Phi^T$.}
\label{figlayerdiagram}
\end{figure}
sketched in figure~\ref{figlayerdiagram}.  To improve computational speed, we truncate values in the residual layer to zero if their absolute value is below 0.005.  In the context of convolutional networks with no downsampling, the term $\Phi \vecy$ corresponds to $\sum_j f_j * \vecz_j$, and we consider $z$ to be an $M$-by-$N$-by-$J$ layer of neurons.  The transpose $\Phi^T$ corresponds to convolutions with the reflections of $f_j$.

To learn the weights~$\Phi$, we use stochastic gradient descent on equation~(\ref{deconvenergy}).  Writing the energy as
\begin{equation}
E = \frac{1}{2}\sum_p\left(x(p) - \sum_q f_j(q)z_j(p-q)\right)^2\,,
\end{equation}
we see that the change in a kernel~$f_j$ at one point~$q$ is given by
\begin{align}
\Delta f_j(q) \propto -\frac{\partial E}{\partial(f_j(q))} &= \sum_p \left(x(p)-\sum_q f_j(q)z_j(p-q)\right)z_j(p-q)\\
&= \sum_p \vecr(p)\vecz_j(p-q).
\label{updaterule}
\end{align}
Note that the weights~$\Phi$ connect the feature maps~$\vecz$ to tho residual layer~$\vecr$ (Figure~\ref{figlayerdiagram}).
This is reminiscent of the Hebbian rule $\Delta\Phi\propto \vecr\vecy^T$.  However, the fact that we are replicating patches across the feature maps means that the  change in a weight is given by the sum over all replications of the kernel.  Equation~(\ref{updaterule}) has the form of a Hebbian rule.  The sum over~$p$ reflects the fact that in the convolution, each weight is repeated over the image domain.  The sum consists of each presynaptic/postsynaptic pair that is connected by the weight~$f_j(q)$.

\section*{Methods}
\label{methods}
We used PetaVision~\cite{petavision}, an open-source neural network simulator that uses OpenMPI for parallel computation.

Our training set was taken from 482 Vine videos posted between Jan.~24 and Jan.~31, 2013.  Each video was converted to a sequence of frames, for 79891 images, which were downsampled from 480x480 to 128x128 using GDAL.  The images were then passed through a center-surround filter with mean~0 and $\sigma$-values of 0.5 for the center and 5.5 for the surround.  For each choice of patch size, number of kernels and stride, we displayed each image for 200 timesteps, using a threshold of $\lambda=0.050$ and updating the weights at the end of the 200 timesteps.  The initial value of the internal state $u_k$ was random for the first frame, and for subsequent images, the initial state was the previous image's final state.  Our training runs used a 64-process parallel computation on an AMD Opteron 6272-based machine.

At the end of the training run, we tested the dictionary by using it to generate sparse representations of the frames of one of the Vine videos, also downsampled to 128x128.  Each image of the frame was shown for 200 timesteps.  The entire video was repeated 10 times to eliminate startup artifacts.  The representations obtained on the second pass were significantly different from that of the first, but the second through tenth passes were all substantially similar to each other.  Although all training runs were performed with a threshold $\lambda=0.050$, reconstructions were run with threshold values of $\lambda=0.025, 0.050, 0.075, 0.100$.  The test runs were performed using a 16 processes on an AMD Opteron 8354-based machine.

We then evaluated the quality of the sparse representations by computing the fraction of V1 cells with zero activity (the percent inactive), and the $l^2$ norm of the residual error, normalized by the $l^2$ norm of the center-surround filtered image (the percent error).  Good representations therefore have percent inactive close to~1.0, and percent error close to zero.

\section*{Results}
\label{results}

\subsection*{Varying strides with fixed number of kernels}
\label{varyscalefactor_fixednfp}

\begin{figure}
\centering
\subfloat[][Illustration of overlaps used in parts (b)-(f).]{\includegraphics[width=0.4\textwidth]{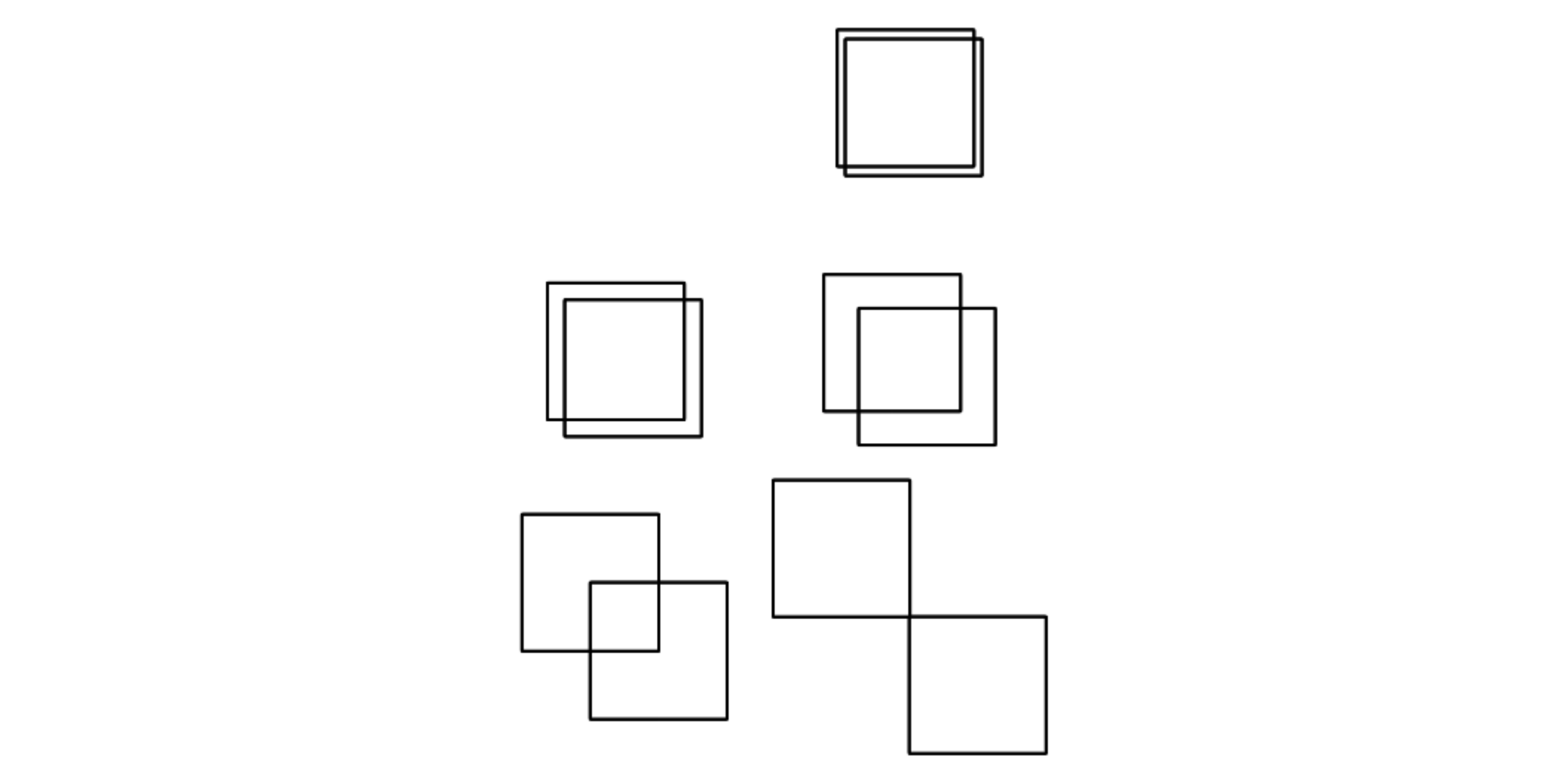}}
\subfloat[][Stride 1, 32x overcomplete.]{\includegraphics[width=0.4\textwidth]{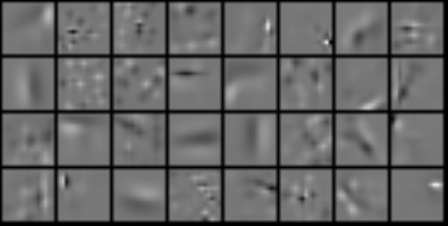}}

\subfloat[][Stride 2, 8x overcomplete]{\includegraphics[width=0.4\textwidth]{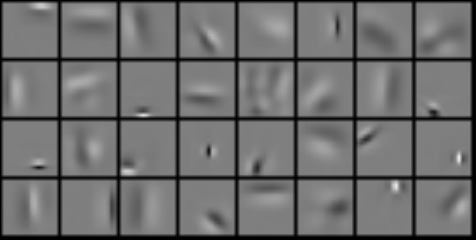}}\quad
\subfloat[][Stride 4, 2x overcomplete]{\includegraphics[width=0.4\textwidth]{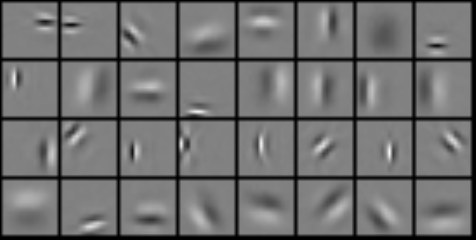}}

\subfloat[][Stride 8, 1/2 undercomplete]{\includegraphics[width=0.4\textwidth]{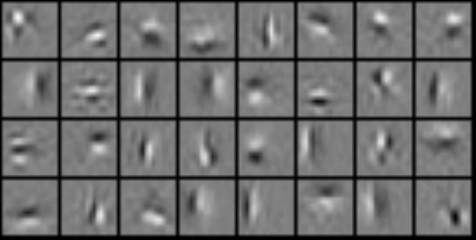}}\quad
\subfloat[][Stride 16, 1/8 undercomplete]{\includegraphics[width=0.4\textwidth]{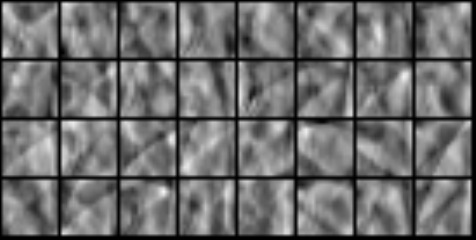}}
\caption{Convolution kernels learned for varying strides with 32 convolution kernels.}
\label{figdictionaries_varyscalefactor_fixednfp}
\end{figure}

We trained several dictionaries using differing strides (1, 2, 4, 8 and 16) and 32 convolutional kernels.  For each run, we used 16x16-pixel patch sizes, except for the stride~1 case, which used 15x15-sized patches.  The reason for this difference is that a V1 neuron's receptive field should be centered on the V1 neuron.  For odd strides, V1 neurons lie above image pixels, but for even strides, each V1 neuron lies above the center of a 2x2-pixel image patch. For stride~$F$, the overcompleteness is $32/F^2$.  Accordingly the problem should be overcomplete for strides 1, 2, and 4; and undercomplete for strides 8 and~16.  In figure~\ref{figdictionaries_varyscalefactor_fixednfp} we show the convolutional kernels learned in each of these runs.  In the most overcomplete case, scale factor~1,  there are several Gabor-like features, as well as several filters with high frequencies.  This is consistent with a highly overcomplete dictionary.  For scale factors 2 and~4, the resulting kernels are primarily Gabor-like.  For a stride of~8, some Gabor-like features are visible but the quality of the kernels is visibly less than for the lower strides.  Finally, for stride~16, the filters do not appear to show a useful dictionary.

\begin{figure}
\subfloat[][]{\includegraphics{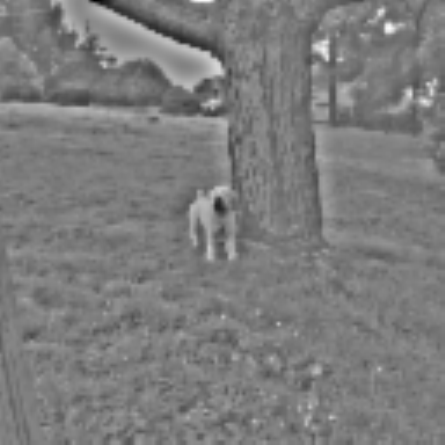}}\quad
\subfloat[][]{\includegraphics{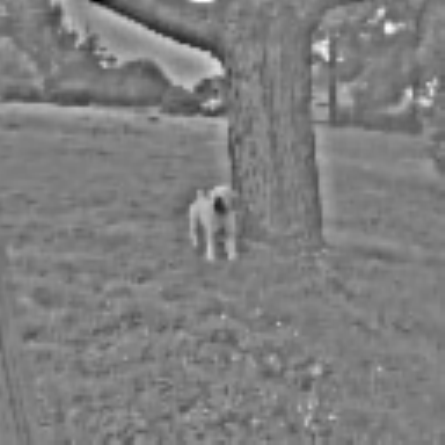}}
\vspace{0.125in}

\subfloat[][]{\includegraphics{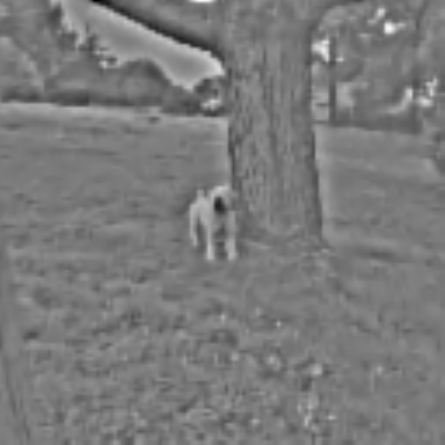}}\quad
\subfloat[][]{\includegraphics{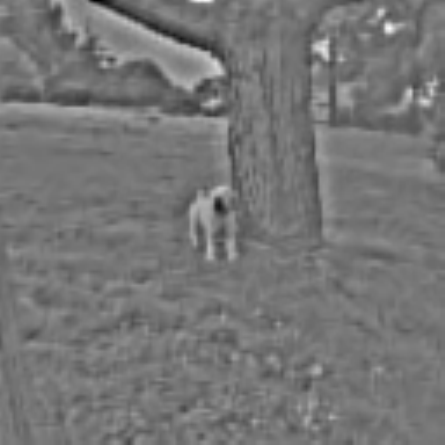}}
\vspace{0.125in}

\subfloat[][]{\includegraphics{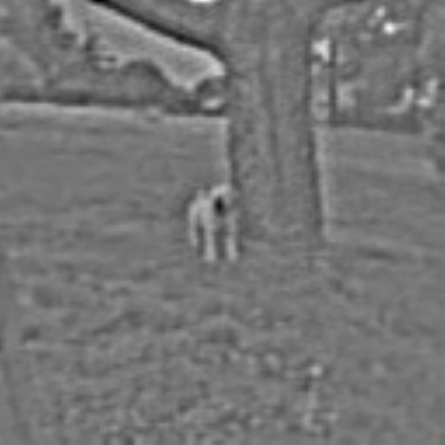}}\quad
\subfloat[][]{\includegraphics{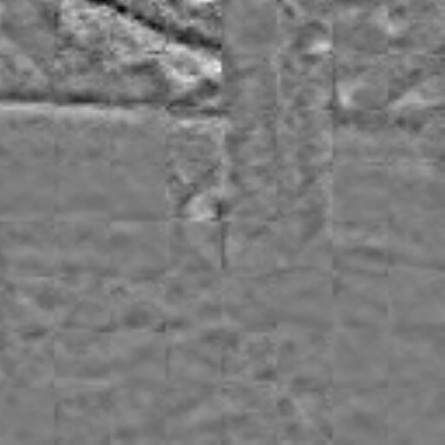}}
\caption{Reconstructions from sparse representations using 32 kernels.  Image (a) is the original image.  Images (b)-(f) were generated using the dictionaries of the corresponding subfigure in figure~\ref{figdictionaries_varyscalefactor_fixednfp}.}
\label{figreconstructions_varyscalefactor_fixednfp}
\end{figure}

Fig~\ref{figreconstructions_varyscalefactor_fixednfp} shows reconstructions from the dictionaries.  Part~(a) shows the original image and (b)-(f) show the reconstructions using the kernels shown in the corresponding parts of figure~\ref{figdictionaries_varyscalefactor_fixednfp}.  Parts (b)-(d), corresponding to overcomplete networks, show good reconstructions.  Part~(e) is slightly undercomplete; the image is recognizable but some loss in fine detail.  Finally in part~(f), the highly undercomplete case, the reconstruction is extremely poor.

\begin{figure}
\includegraphics[width=5in]{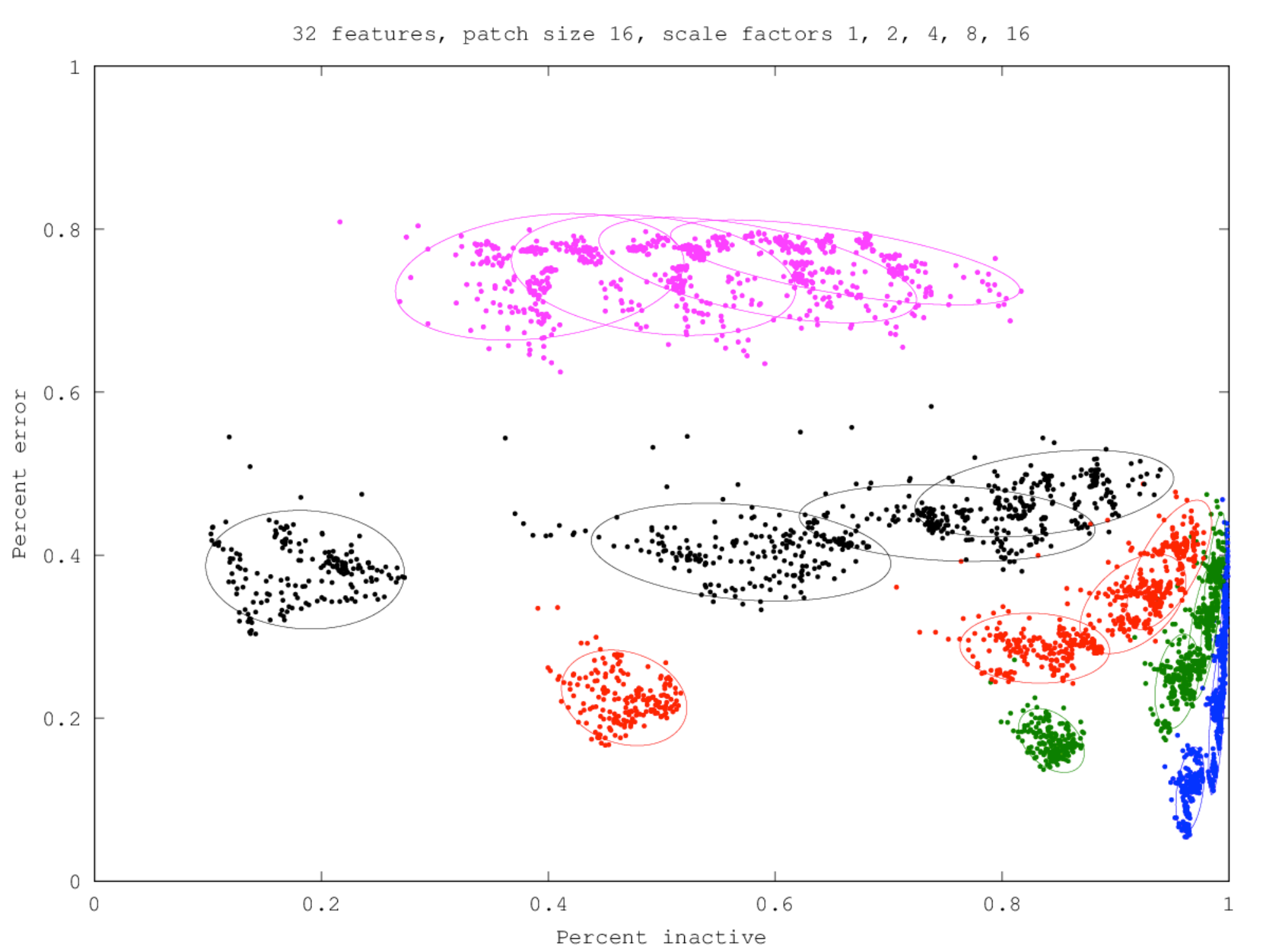}
\caption{Error versus sparsity plots for different strides, using a fixed number of features (32) and fixed patch size (16).  Blue: stride 1, green: stride 2, red: stride 4, black: stride 8, magenta: stride 16.  For each color, there are four runs, corresponding to $\lambda=0.025, 0.050, 0.075, 0.100$.  In each case the percent inactive increases as the threshold increases; generally error increases with increasing threshold.}
\label{figerrorvssparsity_varyscalefactor_fixednfp}
\end{figure}

In figure~\ref{figerrorvssparsity_varyscalefactor_fixednfp}, we show plots of the error versus sparsity.  Twenty runs are depicted: there are five strides (1, 2, 4, 8, and 16), and for each stride there are four thresholds (0.025, 0.050, 0.075, 0.100).  For each run, the 196 frames of the video are shown as a point cloud, and the 2-$\sigma$ uncertainty ellipse is shown for that point cloud.  Different scale factors are shown with different colors: blue for stride 1 runs, green for stride 2, red for stride 4, black for stride 8, and magenta for stride 16.  Within a color, the lower thresholds have a lower percent inactive, and generally lower percent error.  As the stride shrinks and the amount of overlap increases, we obtain greater and greater overcompleteness, and hence sparser reconstructions with lower errors.

\subsection*{Varying strides with fixed overcompleteness factor}
\label{varyscalefactor_fixedovercompleteness}

\begin{figure}
\subfloat[][Illustration of overlaps used in parts (b)-(f).]{\includegraphics[width=0.4\textwidth]{overlap.pdf}}\quad
\subfloat[][Stride 1, 2 kernels.]{\includegraphics[width=0.4\textwidth]{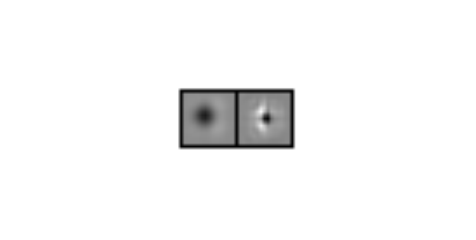}}

\subfloat[][Stride 2, 8 kernels.]{\includegraphics[width=0.4\textwidth]{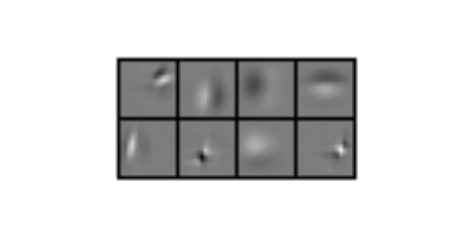}}\quad
\subfloat[][Stride 4, 32 kernels.]{\includegraphics[width=0.4\textwidth]{dictionaries_ReconLCA16x16ScaleFactor4_nfp32_Threshold_0_050.pdf}}
\vspace{0.1in}

\subfloat[][Stride 8, 128 kernels.]{\includegraphics[width=0.4\textwidth]{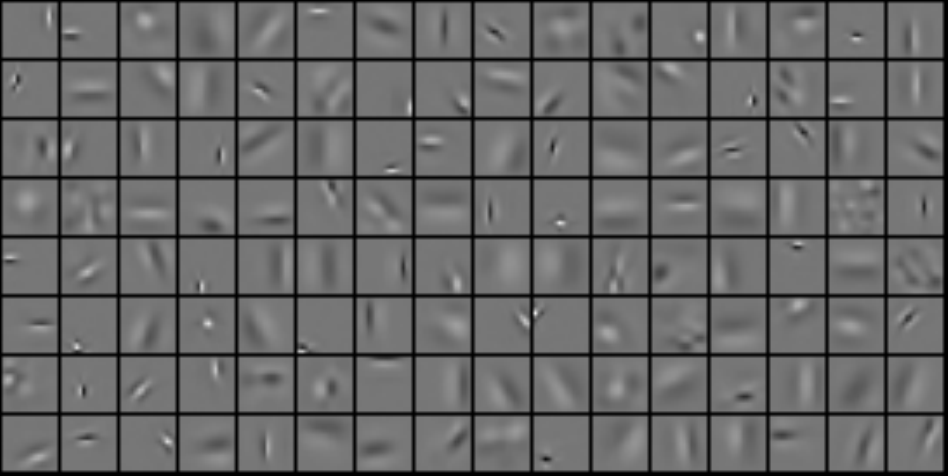}}\quad
\subfloat[][Stride 16, 512 kernels.]{\includegraphics[width=0.4\textwidth]{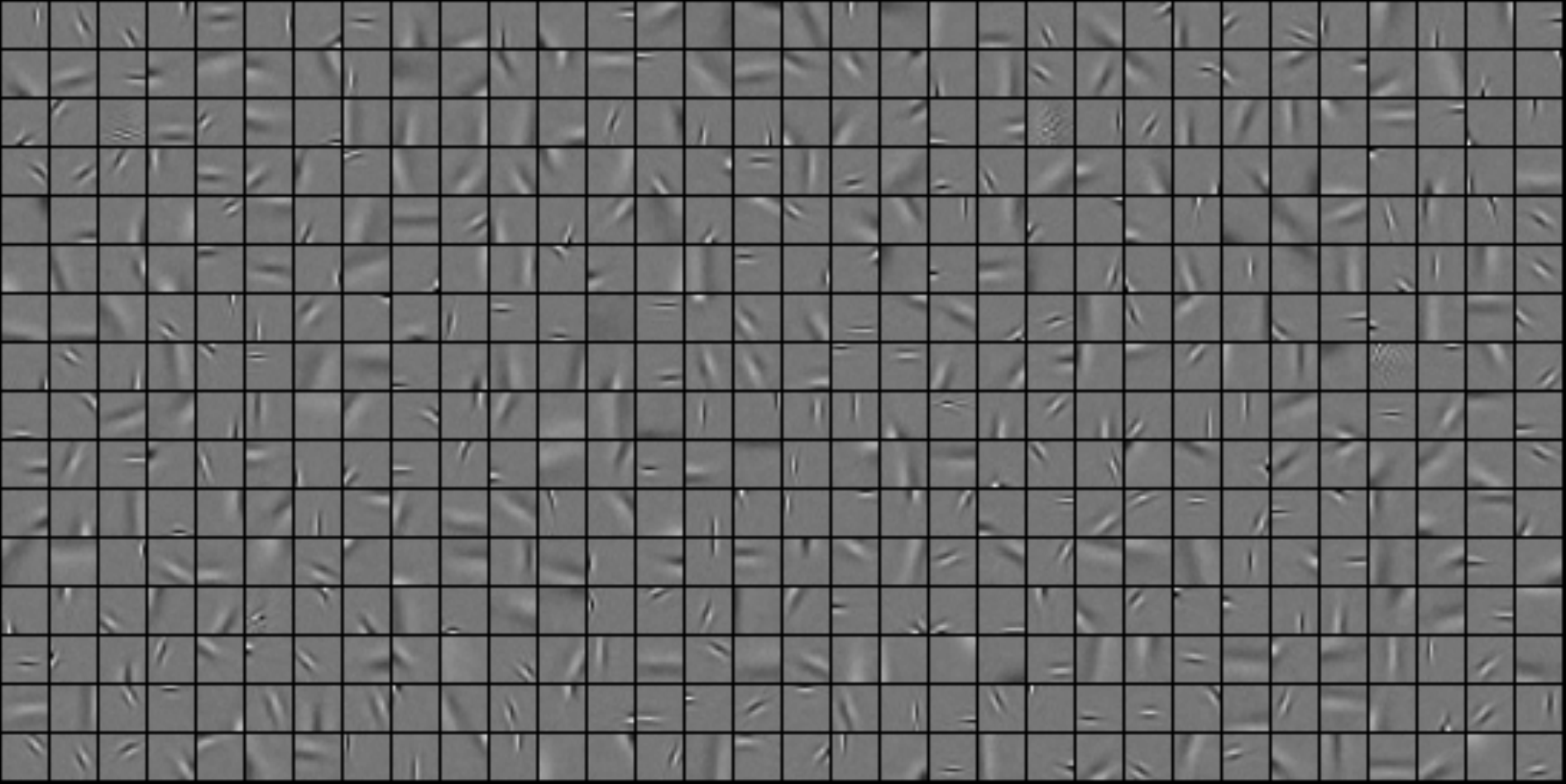}}
\caption{Convolution kernels learned for varying strides with 2x overcompleteness.  All patches are the same size, but the patches in (e) and~(f) are shrunk to fit the figure on the page.}
\label{figdictionaries_varyscalefactor_fixedovercompleteness}
\end{figure}

\begin{figure}
\subfloat[][]{\includegraphics{reconstructions_ganglion.pdf}}\quad
\subfloat[][]{\includegraphics{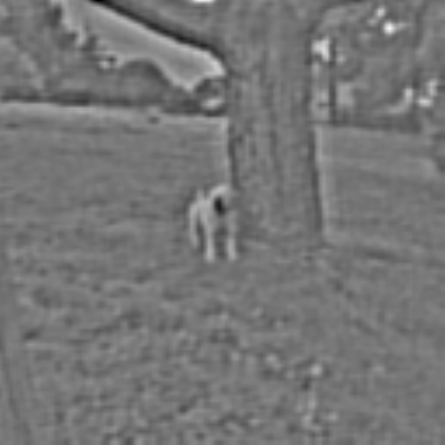}}
\vspace{0.125in}

\subfloat[][]{\includegraphics{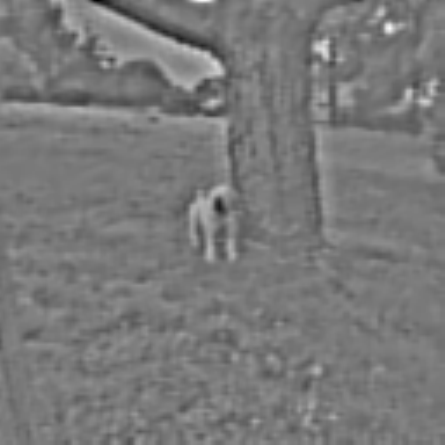}}\quad
\subfloat[][]{\includegraphics{reconstructions_ReconLCA16x16ScaleFactor4_nfp32_Threshold_0_050.pdf}}
\vspace{0.125in}

\subfloat[][]{\includegraphics{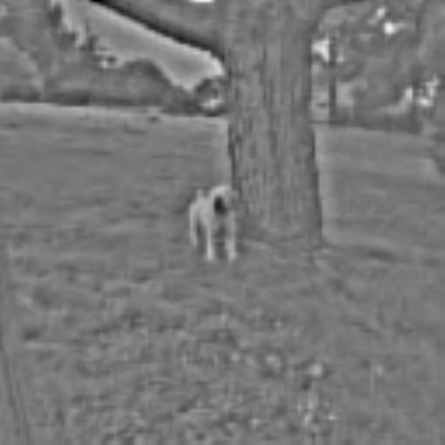}}\quad
\subfloat[][]{\includegraphics{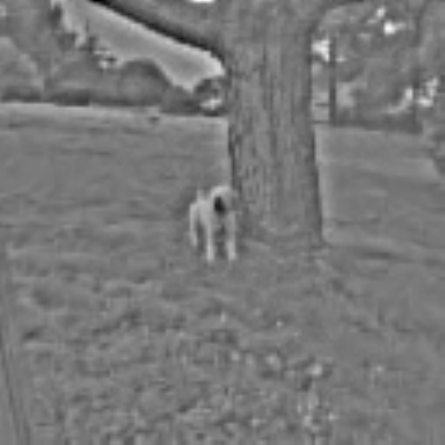}}
\caption{Reconstructions from sparse representations using 2x overcompleteness.  Image (a) is the original image.  Images (b)-(f) were generated using the dictionaries of the corresponding subfigure in figure~\ref{figdictionaries_varyscalefactor_fixedovercompleteness}.}
\label{figreconstructions_varyscalefactor_fixedovercompleteness}
\end{figure}

For the next set of results, we used the same set of strides and the same patch sizes, and chose the number of kernels for each stride so that the dictionary was twice overcomplete.  Thus, for stride 1, we used 2 kernels; for stride 2, 8 kernels, and so forth until for stride 16 (the nonoverlapping case), we need 512 kernels.  In figure~\ref{figdictionaries_varyscalefactor_fixedovercompleteness}, we show the kernels learned.  Figure~\ref{figreconstructions_varyscalefactor_fixedovercompleteness} shows the reconstructions: they are all of approximately equal quality visually, although the higher strides preserve slightly more texture detail.

\begin{figure}
\includegraphics[width=5in]{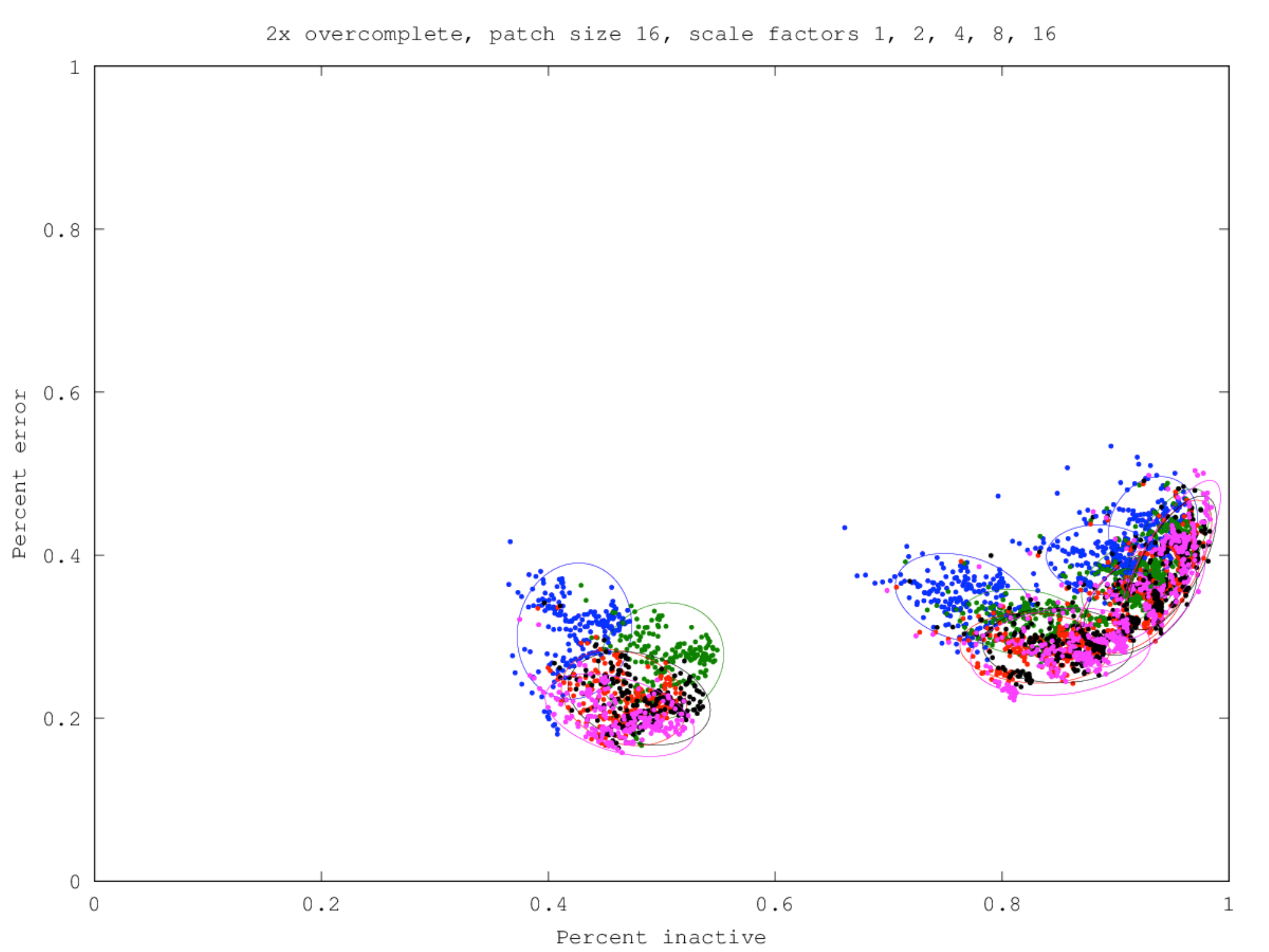}
\caption{Error versus sparsity plots for different strides, using a fixed overcompleteness (2x) and fixed patch size (16x16).  Blue: stride 1, green: stride 2, red: stride 4, black: stride 8, magenta: stride 16.  Values for threshold $\lambda$ are as in figure~\ref{figerrorvssparsity_varyscalefactor_fixednfp}.}
\label{figerrorvssparsity_varyscalefactor_fixedovercompleteness}
\end{figure}

Figure~\ref{figerrorvssparsity_varyscalefactor_fixedovercompleteness} shows the plots of error versus sparsity.  Although the stride-1 case (with only two feature maps) is slightly worse than the other strides, the ellipses for a given threshold show significant overlap.  This result shows that by learning only 2-8 kernels and a small stride of 1 or~2, it is possible to approach the sparse reconstruction produced by 512 kernels that do not overlap.

\subsection*{Varying patch size}
\label{vary_patchsize}

\begin{figure}
\quad
\subfloat[][8x8 patches.]{\includegraphics[width=0.4\textwidth]{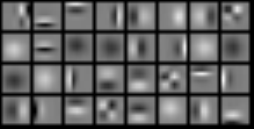}}\quad
\subfloat[][16x16 patches.]{\includegraphics[width=0.4\textwidth]{dictionaries_ReconLCA16x16ScaleFactor2_nfp32_Threshold_0_050.pdf}}
\vspace{0.125in}

\quad
\subfloat[][32x32 patches.]{\includegraphics[width=0.4\textwidth]{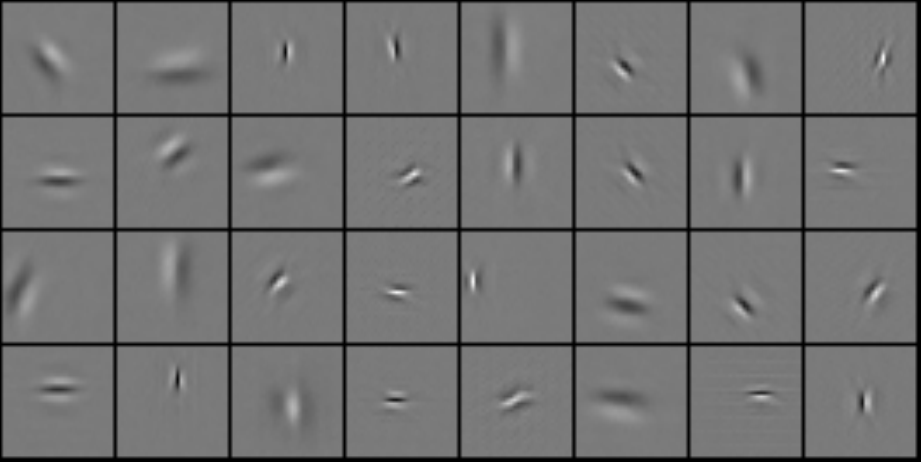}}\quad
\subfloat[][64x64 patches.]{\includegraphics[width=0.4\textwidth]{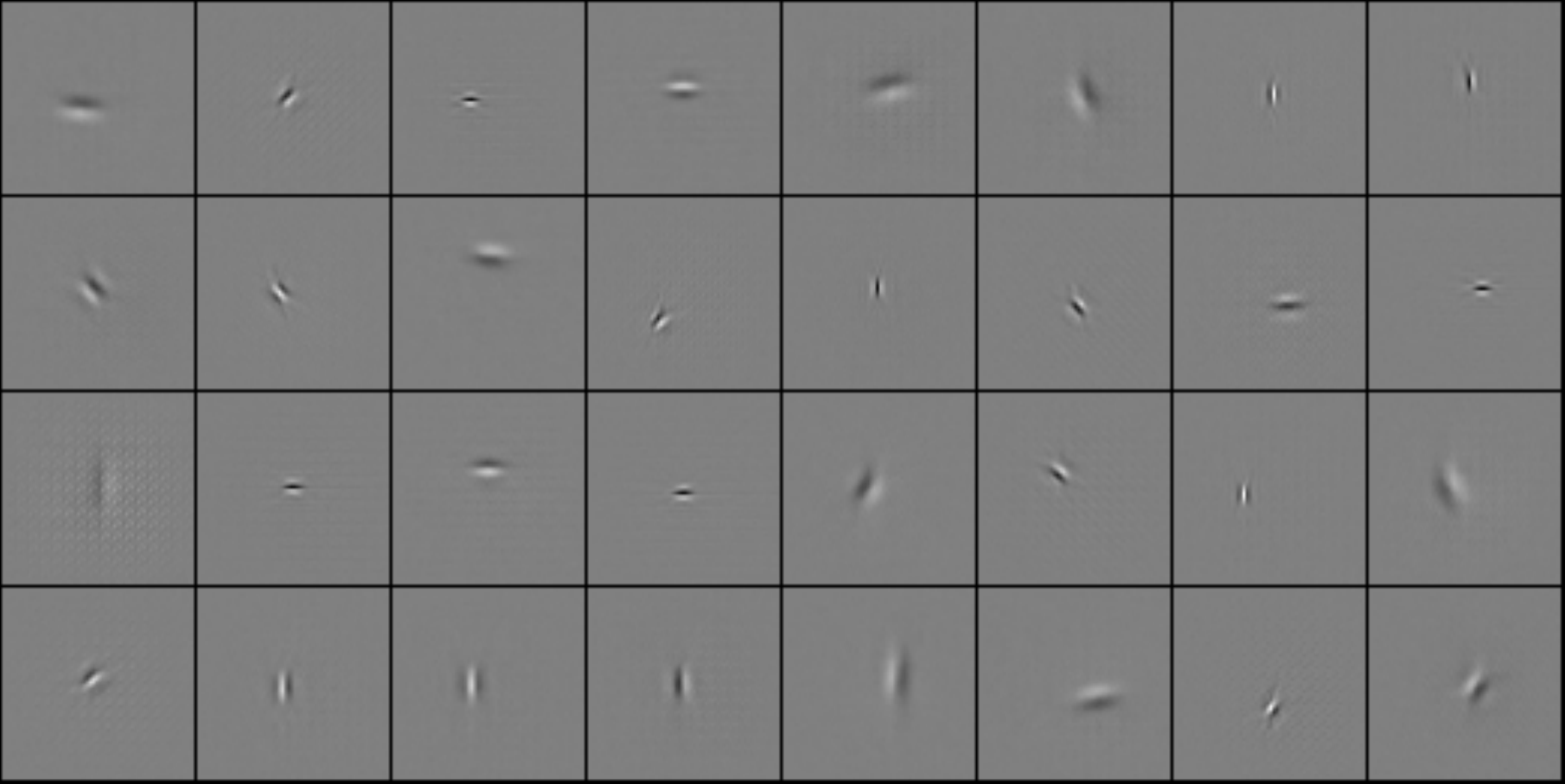}}
\caption{Convolution kernels learned for varying patch sizes with stride~4 and 32 kernels.}
\label{figdictionaries_varypatchsize}
\end{figure}

We previously noted that for fixed stride and number of kernels, overcompleteness should be independent of patch size.  In the final set of results, we test this prediction by measuring the effect of patch size on reconstruction quality.  In figure~\ref{figdictionaries_varypatchsize}, we show the kernels learned using 8x8, 16x16, 32x32, and 64x64 patches.  Although the patch sizes change, we see that the features being learned stay approximately the same size.  Thus, the kernels learned in the 64x64 experiment have a small region of nonzero weights; the larger patch size available does not lead to larger features being learned.

\begin{figure}
\quad
\subfloat[][]{\includegraphics{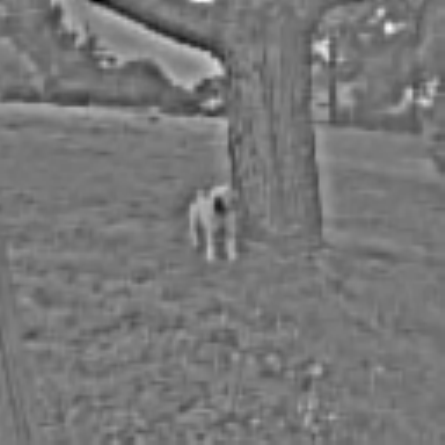}}\quad
\subfloat[][]{\includegraphics{reconstructions_ReconLCA16x16ScaleFactor2_nfp32_Threshold_0_050.pdf}}

\vspace{0.125in}

\quad
\subfloat[][]{\includegraphics{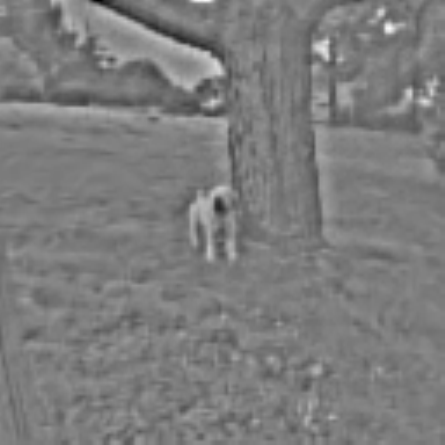}}\quad
\subfloat[][]{\includegraphics{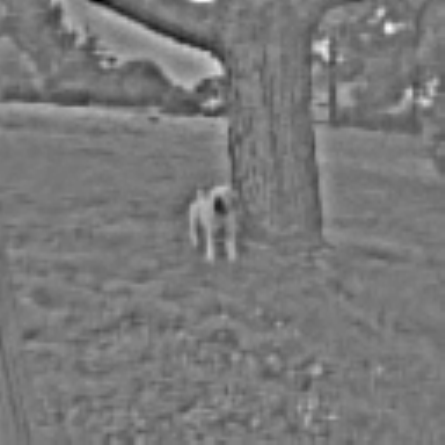}}
\caption{Reconstructions from sparse representations.  Images (A)-(D) were generated using the dictionaries of the corresponding subfigure in figure~\ref{figdictionaries_varypatchsize}.}
\label{figreconstructions_varypatchsize}
\end{figure}

\begin{figure}
\includegraphics[width=5in]{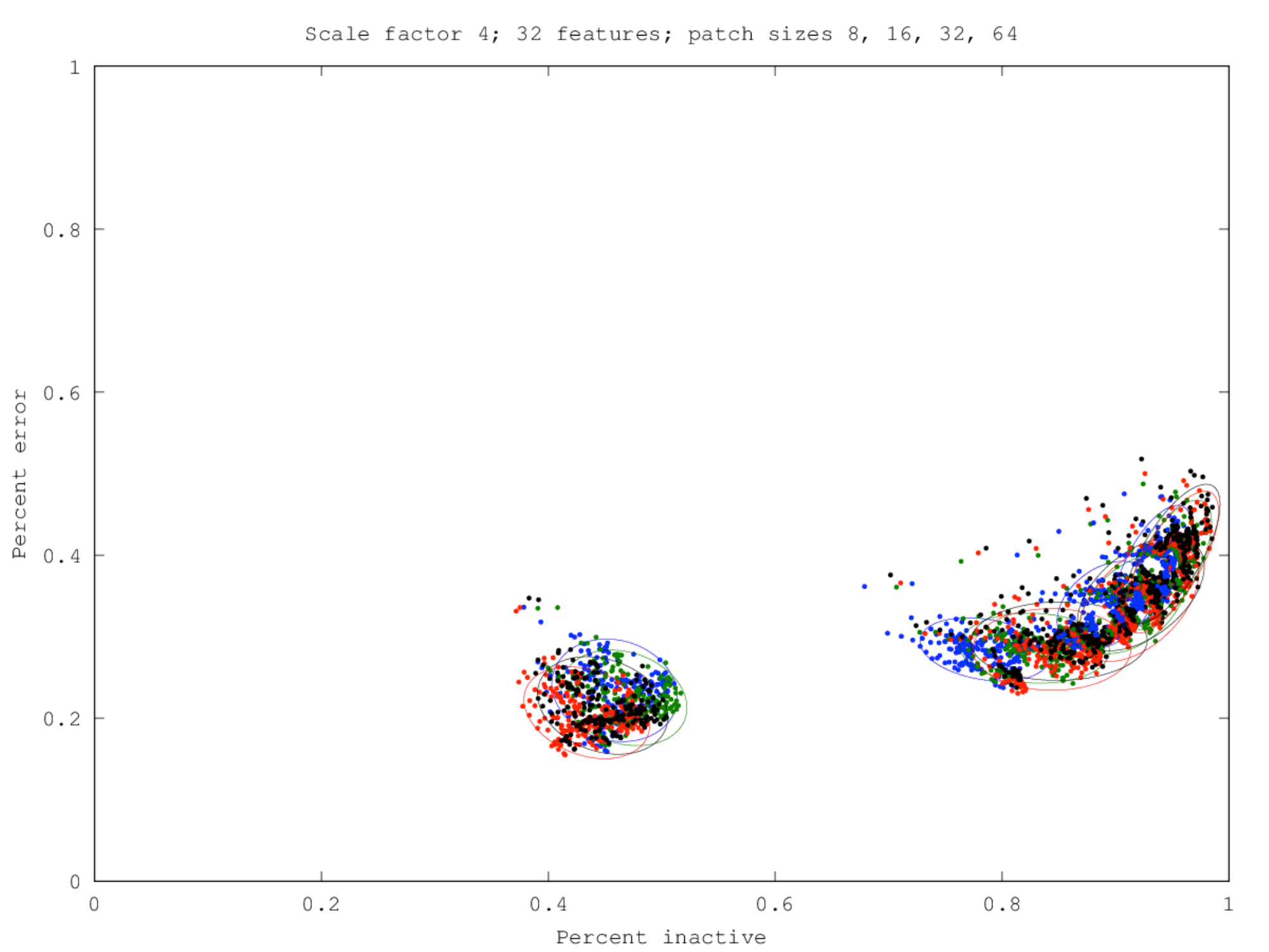}
\caption{Error versus sparsity plots for different patch sizes, using a fixed stride~(4) and fixed number of kernels~(32).  Blue: 8x8, green: 16x16, red: 32x32, black: 64x64.  Values for threshold $\lambda$ are as in figure~\ref{figerrorvssparsity_varyscalefactor_fixednfp}.}
\label{figerrorvssparsity_varypatchsize}
\end{figure}

In Figure~\ref{figreconstructions_varypatchsize}, we show the reconstructions from the kernels shown in figure~\ref{figdictionaries_varypatchsize}.  The visual quality of the reconstruction is similar across all four cases.  This is confirmed in the plot of error versus sparsity in figure~\ref{figerrorvssparsity_varypatchsize}, where we see that for a given threshold, the four reconstructions using different patch sizes have substantially overlapping uncertainty ellipses.  This confirms that over a wide range of patch sizes, the patch size has essentially no effect on overcompleteness or reconstruction quality, given a fixed stride and number of kernels.

\clearpage

\section*{Conclusion}
\label{conclusion}
We have seen that, for a dictionary learned from a deconvolutional network, the overcompleteness and quality of reconstruction is determined by the stride and the number of features, and not the patch size.  Indeed, even for large patches, the learned filters tend to have small regions of support.  Since the number of independent parameters in the dictionary is given by the patch size and the number of features, we observe that we can increase overcompleteness (and hence the quality of the sparse representation) without increasing the number of parameters by increasing the overlap of the receptive fields of adjacent V1 neurons.

\section*{Acknowledgments}

Work performed for the DARPA UPSIDE Program under Cooperative Agreement Award HR0011-13-2-0015.

\bibliographystyle{plain}

\bibliography{replicatingkernels}

\end{document}